\documentclass[reprint,amsmath,amssymb,aps,prl]{revtex4-2}

\usepackage{graphicx}
\usepackage{dcolumn}
\usepackage{bm}
\usepackage{mathrsfs}
\usepackage[utf8]{inputenc}
\usepackage{comment}
\usepackage{xcolor}

\newcommand{\diff}[2]{\frac{\mathrm{d}{#1}}{\mathrm{d}{#2}}}
\newcommand{\piff}[2]{\frac{\partial{#1}}{\partial{#2}}}
\newcommand{\T}{\mathscr{T}}
\newcommand{\C}{\mathscr{C}}

\begin{document}

\title{A Phase Description of Mutually Coupled Chaotic Oscillators}

\author{Haruma Furukawa}
\email{furukawa.haruma.64n@st.kyoto-u.ac.jp}
\affiliation{%
 Graduate School of Informatics, Kyoto University, Yoshida-Honmachi, Sakyo-ku, Kyoto, Japan
}%
\author{Takashi Imai}
\affiliation{%
 Department of Data Science, Shiga University, 1-1-1, Bamba, Hikone-shi, Shiga, Japan
}%
\author{Toshio Aoyagi}
\affiliation{%
 Graduate School of Informatics, Kyoto University, Yoshida-Honmachi, Sakyo-ku, Kyoto, Japan
}%

\begin{abstract}
The synchronization of rhythms is ubiquitous in both natural and engineered systems, and the demand for data-driven analysis is growing.
When rhythms arise from limit cycles, phase reduction theory shows that their dynamics are universally modeled as coupled phase oscillators under weak coupling.
This simple representation enables direct inference of inter-rhythm coupling functions from measured time‑series data.
However, strongly rhythmic chaos can masquerade as noisy limit cycles.
In such cases, standard estimators still return plausible coupling functions even though a phase‑oscillator model lacks a priori justification.
We therefore extend the phase description to the chaotic oscillators.
Specifically, we derive a closed equation for the phase difference by defining the phase on a Poincar\'e section and averaging the phase dynamics over invariant measures of the induced return maps. 
Numerically, the derived theoretical functions are in close agreement with those inferred from time-series data. Consequently, our results justify the applicability of phase description to coupled chaotic oscillators and show that data‑driven coupling functions retain clear dynamical meaning in the absence of limit cycles.
\end{abstract}

\maketitle

Synchronization is a ubiquitous nonlinear phenomenon observed across various fields, including neuroscience~\cite{Buzsaki2004, Izhikevich2007, Fell2011}, biology~\cite{Buck1966, Ryan1986, Ermentrout1991, Aihara2009}, and engineering~\cite{Strogatz2005}.
The growing availability of high-resolution experimental and simulation data has fueled a shift toward data-driven approaches in many physical sciences.
A central challenge in this direction is to infer reliable and interpretable models from complex, noisy, and often high-dimensional observations.

Systems that cause rhythm synchronization are commonly modeled as coupled phase oscillators, whose deterministic dynamics are derived by applying the phase reduction theory to limit-cycle oscillator systems~\cite{Winfree1967,Kuramoto1984,Nakao2016}:
\begin{equation}\label{eq:phase-dynamics}
    \dot{\phi}_i = \omega_i + \sum_{j} \Gamma_{ij} \left(\phi_j-\phi_i\right)+\xi_i(t).
\end{equation}
Here, $\phi_i$ and $\omega_i$ denote the phase and the natural frequency of the $i$-th oscillator, and $\Gamma_{ij}(\phi_j-\phi_i)$ is a $2\pi$-periodic function representing the interaction from the $j$-th oscillator to the $i$-th one.
The last term of Eq.~\eqref{eq:phase-dynamics}, $\xi_i(t)$, is introduced phenomenologically to analyze experimental data, representing external disturbances, measurement noise, and model mismatch.
In this framework, phase-coupling functions $\{\Gamma_{ij}\}$ govern the rhythmic collective behavior, hence the crucial practical task is to infer these coupling functions from experimental time-series data.
Recent advances in statistical and machine-learning methods enable the direct inference of these functions~\cite{Tokuda2007,Kralemann2007,Kralemann2008,Revzen2008,Ota2014,Couzin2015,Matsuda2017,Gengel2019,Wodeyar2021,Yawata2024}.

\begin{figure}[t]
    \centering
    \includegraphics[width=0.95\linewidth]{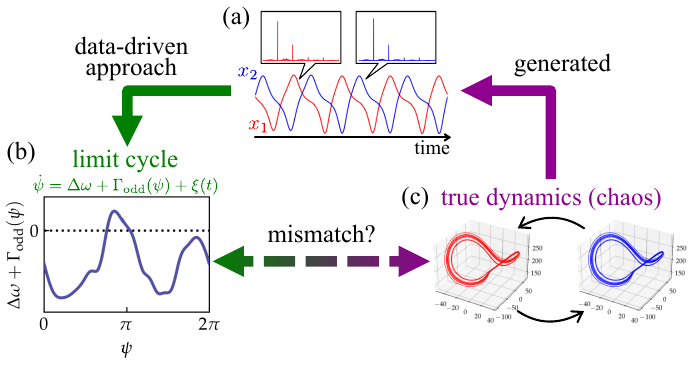}
    \caption{
    Rhythmic time series are often analyzed under the assumption of an underlying limit cycle, but may be generated by rhythmic chaos.
    (a) Nearly periodic signals with sharp spectral peaks appear consistent with the phase description.
    (b) Under this assumption, a standard data-driven procedure infers an effective interaction function from the observed time series (green arrow).
    (c) However, similar rhythmic time series can be generated by a pair of mutually coupled chaotic oscillators (purple arrow). In this case, the limit-cycle hypothesis underlying the inference is violated, making the validity of the inferred coupling function questionable.
}
    \label{fig:intro}
\end{figure}

However, in real experimental settings, a fundamental question arises: does the presence of a rhythm in a time series necessarily imply an underlying limit-cycle oscillator?
To clarify our argument, let us consider the case in which the signals shown in Fig.~\ref{fig:intro}(a) are generated numerically by a two-oscillator system.
The observed signals exhibit features characteristic of perturbed limit-cycle oscillations.
Therefore, we apply a standard statistical procedure that treats the system as coupled phase oscillators.
We define the phase difference $\psi:=\phi_1-\phi_2$ and infer the function $\Gamma_{\rm odd}(\psi)$ that governs rhythmic dynamics, given by
\begin{align}\label{eq:df-phase-dynamics}
    \dot{\psi}=\underbrace{\omega_1-\omega_2}_{=:\Delta\omega}+\underbrace{\Gamma_{12}(-\psi)-\Gamma_{21}(\psi)}_{=:\Gamma_{\rm odd}(\psi)}.
\end{align}
The inferred function is shown in Fig.~\ref{fig:intro}(b). This function provides an average description of the dynamics of $\psi$ and appears to be valid at first glance. In reality, however, there is a fundamental inconsistency: the signals are not generated by limit-cycle oscillators. 
Figure~\ref{fig:intro}(c) displays trajectories of the oscillators used to generate the data, indicating the presence of {\it chaotic} attractors.

In this artificial example, limit-cycle oscillations and chaotic oscillations can be easily distinguished. 
However, in experimental data, distinguishing them often becomes challenging due to the interference of noise.
For instance, whether certain biological rhythms, such as EEG activity or cardiac dynamics, are truly chaotic remains a topic of debate~\cite{Korn2003,Glass2009}.

From a theoretical standpoint, the phase reduction does not apply to chaotic oscillators.
On the other hand, it is widely known that chaotic oscillators also exhibit phase locking~\cite{Lorenz1963,Farmer1980,Rosenblum1996,Pikovsky1997,Boccaletti2002,Osipov2003}, similar to limit-cycle oscillators.
This coexistence of similarity and difference suggests that the phase reduction can be extended to coupled chaotic oscillators, though such an extension has not been achieved.

\begin{figure}
    \centering
    \includegraphics[width=0.95\linewidth]{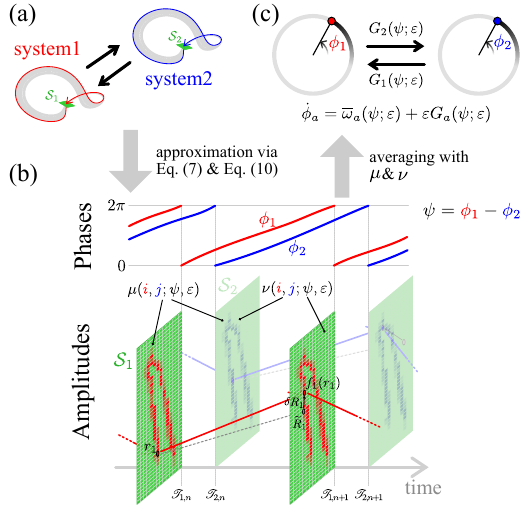}
\caption{Schematic of the proposed theory.
(a) We consider two weakly coupled chaotic oscillators that can exhibit phase synchronization.
(b) For each oscillator, we define a cross section $\mathcal{S}$ and decompose the dynamics into an amplitude $R$ and a phase $\phi$. 
Assuming a time-scale separation where the amplitude dynamics evolve much faster than the phase difference $\psi$, we treat $\psi$ as effectively constant over a single period.
To capture the chaotic nature of the amplitude fluctuations, we construct two invariant measures, $\mu$ and $\nu$, for the induced return maps.
(c) Averaging the phase dynamics with respect to these measures yields the effective phase-difference dynamics.
}
    \label{fig:theory_image}
\end{figure}

In this Letter, we demonstrate that chaotic oscillators admit a phase description at the level of the phase difference. Our approach is based on a recent framework for a single chaotic oscillator under periodic forcing~\cite{Imai2022}. The results of this Letter clarify the existence of a rhythm interaction function between chaotic oscillators, achieving a unified description with limit-cycle oscillators while simultaneously highlighting the unique characteristics of chaotic oscillators not found in limit-cycle oscillators. An overview of our proposed theory is illustrated in Fig.~\ref{fig:theory_image}.


We consider mutually weakly coupled chaotic oscillators:
\begin{eqnarray}\label{eq:coupled original system}
    \begin{cases}
        \dot{X}_1=F_{1}(X_1)+\varepsilon P_1(X_1,X_2), \\
        \dot{X}_2=F_{2}(X_2)+\varepsilon P_2(X_2,X_1),
    \end{cases}
    \quad X_1,X_2 \in \mathbb{R}^N.
\end{eqnarray}
Since chaotic oscillators generally lack the isochrons required for a phase description, we introduce a transverse $(N{-}1)$-dimensional hyperplane (cross section) $\mathcal{S}_a$ for the oscillator $a\in\{1,\,2\}$ as an alternative.
The cross section is chosen so that the return times vary only slightly among trajectories initiated from points on it.
The existence of such a cross section is ensured by the system's strongly rhythmic nature.

Since $\mathcal{S}_a$ is not an isochron, we cannot define a phase in the same manner as for a limit-cycle oscillator. 
This is because the return time to $\mathcal{S}_a$, denoted by $T_a$, depends on the initial point $x \in \mathcal{S}_a$.
One might attempt, in the absence of coupling, to define phase dynamics by $\dot{\phi}(t;x)=2\pi/T_a(x)$ for each initial point $x \in \mathcal{S}_a$.
However, this construction yields no well-defined ``phase response'' $\partial \phi_a/\partial X_a$, which makes perturbative calculations in the coupled system [Eq.~\eqref{eq:coupled original system}] impossible.
To overcome this difficulty, instead of defining phases at all points on $\mathcal{S}_a$, we assign linearly growing phases only at a finite set of points indexed by $i$ on $\mathcal{S}_a$.
The phase in the vicinity of the attractor is then defined by smoothly interpolating between these phases.

The resulting dependence of the phase on the index $i$ is undesirable for a proper phase description.
We therefore average the phase dynamics over two different measures defined on the Cartesian product $\mathcal{S}_1 \times \mathcal{S}_2$.
These measures are constructed as invariant measures of a discrete dynamical system that approximates the time evolution of the non-phase variables of the oscillator, hereafter referred to as the {\it amplitude}.

To derive the discrete dynamics of amplitudes, $\mathcal{S}_a$ is partitioned into disjoint cells $\{s_i^{(a)}\}_{i\in I_a}$, and the phase dynamics are defined from the center of each cell $s_i^{(a)}$ as
\begin{align}\label{eq:phi}
    \dot{\phi}_a(t;i)=\omega_a(i)+\varepsilon \left.\piff{\phi_a}{X_a}\right|_{X_a=X_a(t)}P_a(X_a,X_b).
\end{align}
Here, $\omega_a(i):=2\pi/T_a(i)$ and $T_a(i)$ denotes the return time to $\mathcal{S}_a$ for a trajectory starting from the center of $s_i^{(a)}$.
We assume that the cross section is sufficiently finely partitioned.
Hereafter, the subscripts $a,b \in \{1,2\}$ refer to distinct oscillators.

Since the phase is one-dimensional, the remaining $N{-}1$ degrees of freedom constitute the amplitude $R_a$, whose dynamics are governed by
\begin{align}\label{eq:R}
    \dot{R}_a=F_{a,R}(R_a,\phi_a)+\varepsilon \left.\piff{R_a}{X_a}\right|_{X_a=X_a(t)}P_a(X_a,X_b).
\end{align}

For clarity in the subsequent discussion, we introduce the following notation.
In the uncoupled case ($\varepsilon=0$), the orbit, amplitude, and phase of the oscillator $a$ starting from the center of cell $s_i^{(a)}$ are denoted by $\widetilde{X}_a(t;i)$, $\widetilde{R}_a(t;i)$, and $\widetilde{\phi}_a(t;i)$, respectively.
When the trajectory returns to the cross section for the $n$-th time, let $\T_{a,n}$ denote the time and $\C_{a,n}$ denote the index of the cell to which its state belongs.
For convenience, we write $\widetilde{X}_{a,n}(t) := \widetilde{X}_a(t - \T_{a,n}; \C_{a,n})$.

From Eq.~\eqref{eq:R}, the differential equations governing deviations of the amplitudes from uncoupled ones, $\delta R_a(t):=R_a(t)-\tilde{R}_a(t-\T_{a,n};\C_{a,n})$, is approximated by the following linear differential equation up to $O(\varepsilon)$:
\begin{align}\label{eq:dR}
    \begin{split}
        \diff{}{t}\delta R_a \simeq & \left.\piff{F_{a,R}}{R_a}\right|_{X_a=\widetilde{X}_{a,n}(t)}\delta R_a \\ &+ \varepsilon \left.\piff{R_a}{X_a}\right|_{X_a=\widetilde{X}_{a,n}(t)}P_a(\widetilde{X}_{a,n}(t),\widetilde{X}_{b,m_b}(t)),
    \end{split}
\end{align}
where $m_b$ represents the number of times the trajectory of the oscillator $b$ has returned to the cross section $\mathcal{S}_b$. The solution of Eq.~\eqref{eq:dR} evaluated at $t=\T_{a,n+1}$ is given by
\begin{eqnarray}\label{sol:dRs}
    \delta R_a(\T_{a,n+1})=\Phi_{a,n}(\T_{a,n+1})\left[ \delta R_{a,n}+\varepsilon H_{a,n} \right],
\end{eqnarray}
Here, $\delta R_{a,n}:=R_a(\T_{a,n})-\widetilde{R}_a(0;\C_{a,n})$ and
\begin{align*}
    \Phi_{a,n}(t) &:=\mathcal{T}\exp\left( \int_{\mathscr{T}_{a,n}}^{t}\left.\piff{F_{a,R}}{R_a}\right|_{X=\widetilde{X}_{a,n}(s)}\mathrm{d}s \right), \\
    H_{a,n} &:= \int_{\T_{a,n}}^{\T_{a,n+1}}\Upsilon_{a,n}(t)P_a(\widetilde{X}_{a,n}(t),\widetilde{X}_{b,m_b}(t))\mathrm{d}t, \\
    \Upsilon_{a,n}(t)&:=\Phi_{a,n}^{-1}(t)\left.\piff{R_a}{X_a}\right|_{X=\widetilde{X}_{a,n}(t)},
\end{align*}
where $\mathcal{T}\exp$ denotes the time-ordered exponential.

To compute the integral term $H_{a,n}$, we assume that the oscillators return to the cross sections alternately, i.e., $\cdots<\T_{1,n}<\T_{2,n}<\T_{1,n+1}<\T_{2,n+1}<\cdots$.
Furthermore, by ignoring the fluctuations of order $O(\varepsilon)$ in return time due to perturbations, we change the integration variable from $t$ to $\phi_a$ by setting $\phi_a=\omega_a(\C_{a,n})t$.
Depending on the value of $m_b$, the integration interval $[\T_{a,n}, \T_{a,n+1})$ is divided into two parts, $[\T_{a,n}, \T_{b,n})$ and $[\T_{b,n}, \T_{a,n+1})$.

Since we are interested in the synchronization characteristics of chaotic oscillators, we focus only on the vicinity of the synchronization region and treat the phase difference $\psi$ as a constant during integration.
Taking the above into account, $H_{a,n}$ can be approximated as follows:
\begin{widetext}
\begin{alignat}{1}
\begin{split}
    H_{a,n}
    \approx{}& \frac{1}{\omega_a(\C_{a,n})}\int_{0}^{\xi_a}\Upsilon_{a,n}\left(\frac{\phi_a}{\omega_a(\C_{a,n})};\C_{a,n}\right)P_a\left(\widetilde{X}_a\left(\frac{\phi_a}{\omega_a(\C_{a,n})};\C_{a,n}\right),\widetilde{X}_b\left(\frac{\phi_b}{\omega_b(\C_{b,m_b})};\C_{b,m_b}\right)\right)\mathrm{d}\phi_a \\
    + &\frac{1}{\omega_a(\C_{a,n})}\int_{\xi_a}^{2\pi}\Upsilon_{a,n}\left(\frac{\phi_a}{\omega_a(\C_{a,n})};\C_{a,n}\right)P_a\left(\widetilde{X}_a\left(\frac{\phi_a}{\omega_a(\C_{a,n})};\C_{a,n}\right),\widetilde{X}_b\left(\frac{\phi_b}{\omega_b(\C_{b,m_b+1})};\C_{b,m_b+1}\right)\right)\mathrm{d}\phi_a.
\end{split}
\end{alignat}
\end{widetext}
Here, $\xi_1=\psi$, $\xi_2=2\pi-\psi$, $m_1=n-1$, $m_2=n$, and 
\begin{align*}
    \phi_b =
    \begin{cases}
        \phi_2 + \psi \bmod 2\pi & \text{if } b = 1, \\
        \phi_1 - \psi \bmod 2\pi & \text{if } b = 2.
    \end{cases}
\end{align*}

Using Eq.~\eqref{sol:dRs}, we define a dynamical system that maps the approximated amplitude on $\mathcal{S}_a$ to that in the next return, denoted as $f_a$. The trajectory of $f_a$ is denoted by $\{r_a(n)\}_{n=0}$, i.e.,
\begin{align}\label{eq:R-dynamics} \notag
    r_a(n+1)&=f_a(r_a(n)) \\
    &:=\widetilde{R}_a(T_a(\C_{a,n}),\C_{a,n})+\delta R_a(\T_{a,n+1}).
\end{align}
Note that $\C_{a,n}$ can be determined from $r_a(n)$.

We consider two types of coupling dynamics and construct an invariant measure for each of them. One is a mapping that transfers $(r_1(n),r_2(n))$ to $(r_1(n+1),r_2(n+1))$, and the other is a mapping that transfers $(r_1(n+1),r_2(n))$ to $(r_1(n+2),r_2(n+1))$. We denote the invariant measure for the former as $\mu(i,j;\psi,\varepsilon)$ and that for the latter as $\nu(i,j;\psi,\varepsilon)$. Here, $i$ and $j$ are the indices of the cells in $\mathcal{S}_1$ and $\mathcal{S}_2$, respectively.

The averaging using them is performed as follows. Approximating Eq.~\eqref{eq:phi} up to $O(\varepsilon)$, we obtain
\begin{align}\label{eq:phi_approx}
    \begin{split}
        &\dot{\phi}_a(t) \simeq \omega_a(\C_{a,n})+\varepsilon \zeta_a(\phi_a;\C_{a,n}) \widetilde{P}_a(\phi_a,\phi_b;\C_{a,n},\C_{b,m}), \\
        &\zeta_a(\phi_a;\C_{a,n}) :=\left.\piff{\phi_a}{X_a}\right|_{X_a=\widetilde{X}_{a,n} \left(\phi_a/\omega_a(\C_{a,n})\right)}.
    \end{split}
\end{align}
Here,
\begin{align*}
    &\widetilde{P}_a(\phi_a,\phi_b;\C_{a,n},\C_{b,m}) \\
    :=&P_a\left(\widetilde{X}_{a,n}(\phi_a/\omega_a(\C_{a,n})),\widetilde{X}_{b,m}(\phi_b/\omega_b(\C_{b,m}))\right).
\end{align*}
We eliminate the dependence on the cells in Eq.~\eqref{eq:phi_approx} by averaging with invariant measures, which yields an equation of the form
\begin{eqnarray}\label{eq: main_result}
        \dot{\psi} = \Delta\omega(\psi;\varepsilon)+\varepsilon[G_1(\psi;\varepsilon)-G_2(\psi;\varepsilon)].
\end{eqnarray}
Here,
\begin{eqnarray}\label{eq:averaging}
    \begin{cases}\displaystyle
        \Delta\omega(\psi;\varepsilon) := \overline{\omega}_1(\psi;\varepsilon)-\overline{\omega}_2(\psi;\varepsilon), \\ \displaystyle
        \overline{\omega}_{a}(\psi;\varepsilon)=\frac{2\pi}{\overline{T}_{a}(\psi;\varepsilon)}, \\ \displaystyle
        G_{a}(\psi;\varepsilon)=\frac{1}{2\pi}\int_{0}^{2\pi}\overline{g}_{a}(\phi_a,\psi;\varepsilon)\mathrm{d}\phi_a,
\end{cases}
\end{eqnarray}
and $\overline{T}_{a}(\psi;\varepsilon)$ denotes the averaged return time to the cross section of the oscillator $a$, and $\overline{g}_{a}(\phi_a,\psi;\varepsilon)$ is defined as the average of the response of the oscillator $a$ to perturbations $\zeta_a\widetilde{P}_a$.
This quantity is calculated using the invariant measures constructed above, but there is a subtle issue dependent on $\phi_a$ and $\psi$ regarding which measure should be used.
Details are provided in the Supplemental Material.


\begin{figure}
    \centering
    \includegraphics[width=0.95\linewidth]{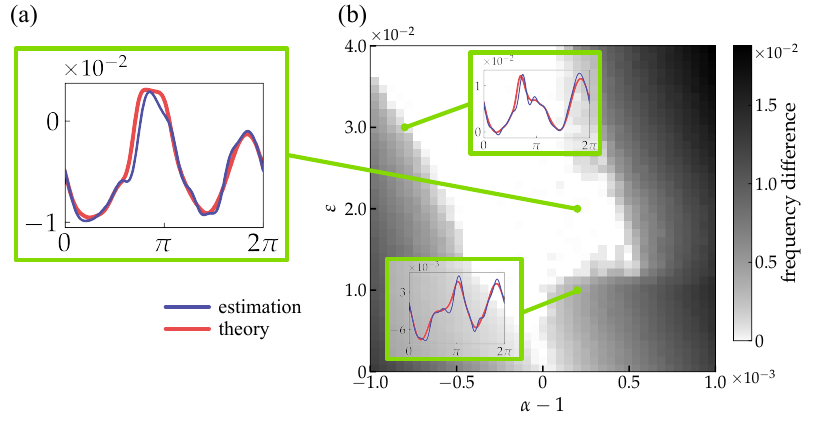}
    \caption{(a) A comparison of the theoretically derived interaction function (red line) and the statistically inferred coupling function (blue line) for the coupled Lorenz system [Eq.~\eqref{eq:double lorenz}]. The parameters are set to $(\alpha,\varepsilon)=(1.0002,0.02)$. The inferred function corresponds to the one shown in Fig.~\ref{fig:intro}(b). (b) The grayscale map illustrates the frequency difference on the $(\alpha,\varepsilon)$ plane, where the white region indicates the Arnold tongue. The insets confirm that our theory accurately predicts the coupling functions even near the chaos-specific distortion of the Arnold tongue.}
    \label{fig:result_lorenz}
\end{figure}

To validate our theory, we employ statistical coupling-function inference as an independent numerical check.
We analyze time series generated by coupled chaotic oscillators without assuming prior knowledge of their chaotic nature, instead adopting the standard limit-cycle framework as a starting point.
By postulating a phase-oscillator form for the phase difference, we infer the coupling function directly from the phase time series as in Fig.~\ref{fig:intro}(a,b). If our theory is correct, this empirically inferred function should coincide with the interaction function predicted by our reduction, even for chaotic systems.
Such agreement would provide a direct test of whether the phase-difference dynamics are effectively captured by our theoretical framework.

As a first example, we revisit the coupled Lorenz system in Fig.~\ref{fig:intro}(c).
This system~\cite{Lorenz1963} is governed by 
\begin{eqnarray}\label{eq:double lorenz}
    \begin{cases}
    \dot{x}_1=\sigma(y_1-x_1), \\
    \dot{y}_1= x_1(\rho-z_1)-y_1+\varepsilon (1+0.01x_2^2), \\
    \dot{z}_1= x_1y_1-\beta z_1, \\
    \dot{x}_2=\alpha[\sigma(y_2-x_2)]+\varepsilon (-1+0.01x_1^2), \\
    \dot{y}_2= \alpha[x_2(\rho-z_2)-y_2], \\
    \dot{z}_2= \alpha[x_2y_2-\beta z_2]-\varepsilon (2+0.01y_1^2).
    \end{cases}
\end{eqnarray}
with parameters $(\rho, \sigma, \beta)=(210, 10, 8/3)$. Here,  $\alpha$ controls the mismatch between the oscillators ($\alpha=1$ for identical oscillators), and $\varepsilon$ is the coupling strength. We set the cross section at $z=\rho-1$ for both oscillators, which yields small variability in return times and is consistent with the assumptions of the theory. 
The phase $\phi$ is reconstructed by assigning  $2n\pi$ at the $n$-th crossing of the section, followed by linear interpolation. The coupling function is then inferred from the resulting phase time series (see Supplemental Material for details).

The results of this inference-based test are summarized in Fig.~\ref{fig:result_lorenz}. As shown in Fig.~\ref{fig:result_lorenz}(a), the coupling function inferred from the time series shows excellent agreement with our theoretical prediction. This consistency holds across various parameter sets [Fig.~\ref{fig:result_lorenz}(b), insets]. Figure ~\ref{fig:result_lorenz}(b) further displays the frequency difference across the $(\alpha,\varepsilon)$ plane, where the white region identifies the 1:1 synchronization regime (Arnold tongue). Notably, for $0.01<\varepsilon<0.03$, the Arnold tongue exhibits a distinct distortion. In chaotic systems, even weak coupling can perturb the underlying trajectories and, consequently, the invariant measures of the induced return maps. This measure dependence modifies the averaged frequency and interaction terms, leading to the observed distortion of the synchronization boundary. Our theory successfully reproduces this effect, which is fundamentally absent in the standard phase reduction for limit-cycle oscillators.

\begin{figure}
    \centering
    \includegraphics[width=0.95\linewidth]{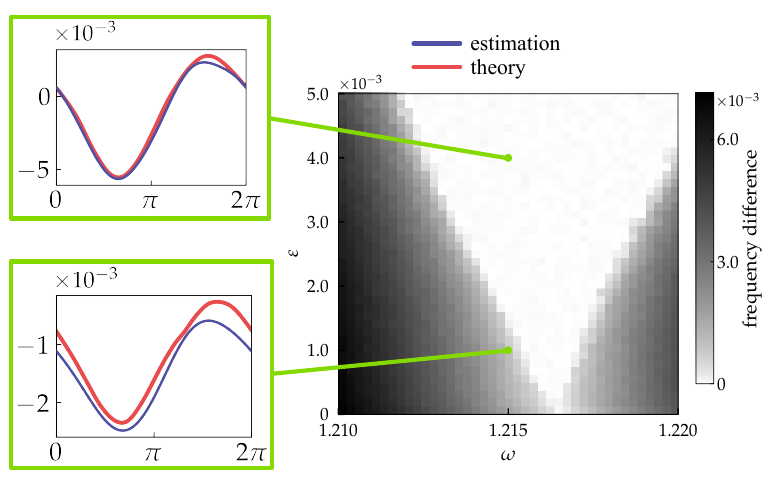}
    \caption{Comparison results for the heterogeneous system composed of R\"{o}ssler and Sprott--N oscillators [Eq.~\eqref{eq:rossler sprottn}]. Same as Fig.~\ref{fig:result_lorenz}, but for the heterogeneous case. Theoretically derived (red) and statistically inferred (blue) coupling functions. (b) The grayscale map representing the frequency difference. The insets demonstrate the agreement between theory and inference for representative parameter sets both inside and outside the synchronization region.} 
    \label{fig:result_rosspr}
\end{figure}

To further demonstrate the broad applicability of our approach, we consider a heterogeneous pair consisting of a R\"{o}ssler  oscillator~\cite{Rossler1976} and a Sprott-N oscillator~\cite{Sprott1994} as follows
\begin{eqnarray}\label{eq:rossler sprottn}
    \begin{cases}
    \dot{x}_1=-\omega y_1-z_1+\varepsilon x_2, \\
    \dot{y}_1=\omega x_1 + 0.2 y_1, \\
    \dot{z}_1=0.2+z_1(x_1-7), \\
    \dot{x}_2=-2y_2, \\
    \dot{y}_2=x_2 + z_2^2, \\
    \dot{z}_2=1+y_2-2z_2+\varepsilon x_1,
    \end{cases}
\end{eqnarray}
where $\omega$ controls the natural frequency of the R\"{o}ssler oscillator.
We vary the natural frequency $\omega$ of the Rössler oscillator to achieve 1:1 phase locking while keeping the Sprott–N oscillator fixed. 
In this case, the choice of the cross section is critical due to the distinct geometries of the attractors. 
In such systems, simple planar sections lead to large variability in return times. 
To minimize return-time variability, we employ data-driven optimal isophases ~\cite{Schwabedal2012} as Poincaré sections (see Supplemental Material). 
As shown in Fig.~\ref{fig:result_rosspr}, our theoretical predictions remain in quantitative agreement with the inferred coupling functions for both synchronized and unsynchronized states. These results demonstrate that the proposed theory is applicable to general heterogeneous chaotic oscillators by selecting appropriate cross sections.

Finally, we summarize the assumptions and limitations of the proposed reduction. We consider two mutually weakly coupled chaotic oscillators.
This approach requires cross sections with a small variability in return times and assumes that the induced return maps have invariant measures, so that averaging over these measures defines the frequency and interaction terms in the phase-difference dynamics.
We focus on near 1:1 synchronization, where crossings of the two sections occur alternately and the phase difference changes little between successive returns.
This separation of time scales justifies treating the phase difference as a constant within the integrals used to compute the interaction terms. 
Extensions to larger networks are conceptually straightforward but become computationally expensive.
In networks of three or more oscillators, the effective interaction terms may depend not only on the phases of a given oscillator pair but also on the phases of other oscillators.
Developing a reduction that systematically captures this network-induced phase dependence is left for future work.

This work was supported by the following: MEXT KAKENHI Grant Numbers JP23H04467; JSPS KAKENHI Grant Numbers JP24H00723, JP20K20520; JST BOOST Grant Number JPMJBS2407.

\bibliographystyle{apsrev4-2}
\bibliography{myref}

\clearpage
\onecolumngrid

\section{supplemental material}
\subsection{Averaging of Phase Dynamics with the Invariant Measures}

In this section, we explain how to average the cell-dependent phase equation [Eq.~\eqref{eq:phi_approx} in the main text] using two types of invariant measures, $\mu$ and $\nu$. 

First, we define the marginal measures as
\begin{equation}
    \mu_{1}(i;\psi,\varepsilon) := \sum_{j}\mu(i,j;\psi,\varepsilon),\qquad \mu_{2}(j;\psi,\varepsilon) := \sum_{i}\mu(i,j;\psi,\varepsilon).
\end{equation}
The measure $\mu_{a}$ can be regarded as an invariant measure of $f_a$, which is defined in Eq.~\eqref{eq:R-dynamics}. We can compute the average return time $\overline{T}_{a}$ of the oscillator $a$:
\begin{equation}
    \overline{T}_{a}(\psi;\varepsilon)=\sum_{i}\mu_{a}(i;\psi,\varepsilon)T_{a}(i).
\end{equation}

Next, we consider the averaged response of the oscillator $a$ to perturbations, denoted by $\overline{g}_{a}(\phi_a,\psi;\varepsilon)$.
Since we suppose the oscillators return to the cross sections alternately, for time $t$ satisfying $\T_{1,n} \leq t < \T_{2,n}$, the oscillator 1 has returned $n$ times to $\mathcal{S}_1$, while oscillator 2 has returned $n{-}1$ times to $\mathcal{S}_2$. In this case, it is appropriate to use the measure $\nu$ to compute the average. Conversely, for $\T_{2,n} \leq t < \T_{1,n+1}$, both the oscillators have returned $n$ times to their respective cross sections, and the measure $\mu$ should be used.

Based on these considerations, we compute $\overline{g}_{a}(\phi_a,\psi;\varepsilon)$ as follows:
\begin{eqnarray}
    \overline{g}_{1}(\phi_1,\psi;\varepsilon) &:=
\begin{cases} \displaystyle
\sum_{i,j}\nu(i,j;\psi,\varepsilon)\left[\frac{T_{1}(i)}{\overline{T}_{1}(\psi;\varepsilon)}\zeta_{1}(\phi_1;i)\right]\cdot\left[\frac{T_{2}(j)}{\overline{T}_{2}(\psi;\varepsilon)}\widetilde{P}_{1}(\phi_1,\phi_1-\psi+2\pi;i,j)\right] & \text{ if $\phi_1-\psi < 0$,} \\ \displaystyle
\sum_{i,j}\mu(i,j;\psi,\varepsilon)\left[\frac{T_{1}(i)}{\overline{T}_{1}(\psi;\varepsilon)}\zeta_{1}(\phi_1;i)\right]\cdot\left[\frac{T_{2}(j)}{\overline{T}_{2}(\psi;\varepsilon)}\widetilde{P}_{1}(\phi_1,\phi_1-\psi;i,j)\right]  & \text{ if $\phi_1-\psi\geq 0$,}
\end{cases} \\
\overline{g}_{2}(\phi_2,\psi;\varepsilon) &:=
\begin{cases} \displaystyle
\sum_{i,j}\mu(i,j;\psi,\varepsilon)\left[\frac{T_{2}(j)}{\overline{T}_{2}(\psi;\varepsilon)}\zeta_{2}(\phi_2;j)\right]\cdot\left[\frac{T_{1}(i)}{\overline{T}_{1}(\psi;\varepsilon)}\widetilde{P}_{2}(\phi_2,\phi_2+\psi;i,j)\right] & \text{ if $\phi_2+\psi\leq 2\pi$,} \\ \displaystyle
\sum_{i,j}\nu(i,j;\psi,\varepsilon)\left[\frac{T_{2}(j)}{\overline{T}_{2}(\psi;\varepsilon)}\zeta_{2}(\phi_2;j)\right]\cdot\left[\frac{T_{1}(i)}{\overline{T}_{1}(\psi;\varepsilon)}\widetilde{P}_{2}(\phi_2,\phi_2+\psi-2\pi;i,j)\right] & \text{ if $\phi_2+\psi > 2\pi$}.
\end{cases}
\end{eqnarray}

In approximating the amplitude dynamics, the phase difference $\psi$ is assumed to be constant. However, strictly speaking, it undergoes slight fluctuations.
In order to take into account their effects, we apply a moving average to Eq.~\eqref{eq: main_result} in the final step.

\subsection{Selected Cross Sections and Trajectories of the Oscillators}

\begin{figure*}
    \centering
    \includegraphics[width=0.95\linewidth]{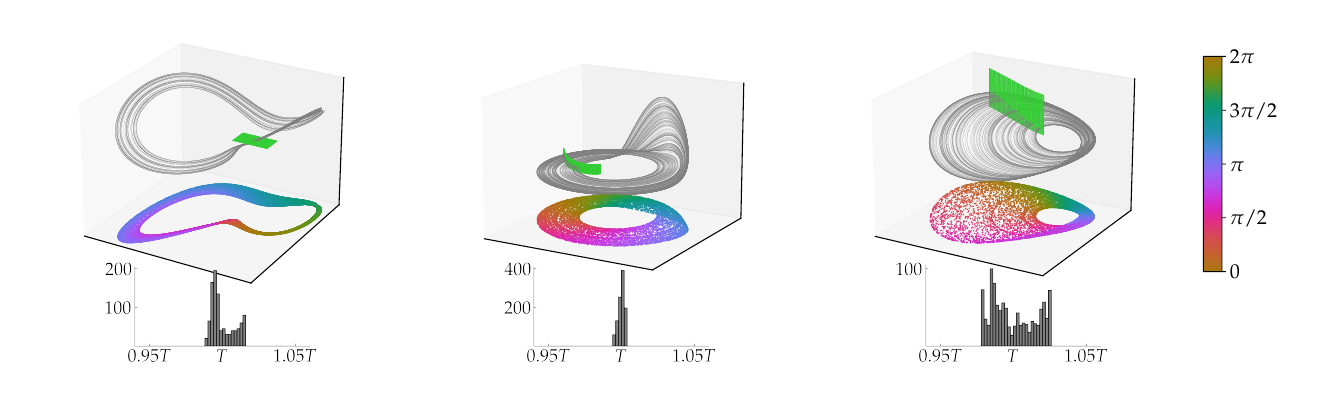}
    \caption{The trajectories of the oscillators (gray), the selected cross sections (green), and the phases determined by the cross sections (the 2D projection shown below each oscillator). The lower panels display the return time histograms for the selected cross sections. For each histogram, the scale $T$ of the horizontal axis represents the average return time to the cross section. The vertical axis shows the density. From left to right: the Lorenz oscillator, the R\"{o}ssler oscillator, and the Sprott-N oscillator.}
    \label{sfig:traj and hist}
\end{figure*}

In this section, we describe the selected cross sections and the method of selection.

For the Lorenz oscillator, the trajectory and the selected cross section ($z=\rho-1$) are shown in the left of Fig.~\ref{sfig:traj and hist}.
Below the trajectory, the phase induced from the cross section is displayed using a color map.
The histogram of return times is also displayed, which confirms that a key assumption for our theory---small variability in return times---is satisfied.

Similar figures are also drawn for the R\"{o}ssler oscillator (center) and the Sprott-N oscillator (right) in Fig.~\ref{sfig:traj and hist}.
For these two oscillators, careful consideration is required in selecting the cross sections.
In fact, selecting a simple plane causes large variations in the return times to the cross section.
This stems from the fact that, unlike the Lorenz oscillator, the amplitude fluctuations in these oscillators are large.
In such cases, it is necessary to use a curved surface rather than a flat one.

To construct the cross section we desire, optimal isophases~\cite{Schwabedal2012} are employed. 
We consider the dynamics of a chaotic oscillator described by $\dot{X}=F(X)$, where $X=(x,y,z)\in\mathbb{R}^3$.  
First, the mean period $T_{\rm mean}$ is numerically estimated by performing a long-time simulation. 
Next, we construct the stroboscopic set $\{X(kT_{\rm mean})\ |\ k=0,1,\cdots,K\}$ and compute the corresponding geometric angle on $xy$ plane at each point.
Defining $r:=\sqrt{x^2+y^2}$, which represents the Euclidean distance from the origin on the $xy$ plane, we perform a linear regression using $\log r$, $(\log r)^2$, and $(\log r)^3$ as independent variables and the geometric angle as the dependent variable. 
Using the obtained optimal weights $w_1,w_2,w_3$, the function $I(r)=w_1\log r + w_2(\log r)^2 + w_3(\log r)^3$ defines the optimal isophase.

For each oscillator, after discarding an initial transient of $1{,}000$ time units, the subsequent $5{,}000$ time units are used for determining the cross section. Numerical integration is carried out using a fourth-order Runge--Kutta method with time step $h=10^{-2}$.

\subsection{Details of Numerical Calculations for the Phase Description}
In this section, we explain in detail the parameters used in our proposed phase description. Specific values are summarized in Table~\ref{stab:sim-params}.

\begin{table}
  \caption{Numerical parameters used in the simulations of the three oscillators.}
  \label{stab:sim-params}
  \begin{ruledtabular}
    \begin{tabular}{lcccc}
      System & Time step $h$ & mesh size & $\Delta R_a$ & $\Delta X_a$ \\
      \hline
      Lorenz & $10^{-3}$ & $(x,y)=(0.02,0.02)$ & $10^{-2}$ & $10^{-2}$ \\
      R\"{o}ssler & $10^{-2}$ & $(r,z)=(0.1,0.0002)$ & $10^{-3}$ & $10^{-3}$ \\
      Sprott-N & $10^{-2}$ & $(r,z)=(0.1,0.02)$ & $10^{-2}$ & $10^{-2}$
    \end{tabular}
  \end{ruledtabular}
\end{table}

To apply our theory, the mesh size must first be determined.
It should be set small enough to capture the details of the oscillator's trajectory in the invariant measure.
The derivatives used in the phase description, $\partial F_a/\partial R_a$ and $\partial R_a/\partial X_a$, are approximated by numerical differentiation with a finite increment $\Delta R_a$ and $\Delta X_a$, respectively.

In our theory, the invariant measures are computed for a fixed phase difference.
The phase difference is fixed at points dividing the interval $[0, 2\pi]$ into 200 equal-length segments.

When inferring invariant measures, the initial state of the constructed discrete dynamics is defined as the center of the cross section, and the first $1{,}000$ steps are removed as transient. The subsequent $30{,}000$ steps are used to construct the empirical distributions, which are then adopted as the invariant measures.

\subsection{Prediction of the Arnold Tongues}
\begin{figure*}
    \centering
    \includegraphics[width=0.95\linewidth]{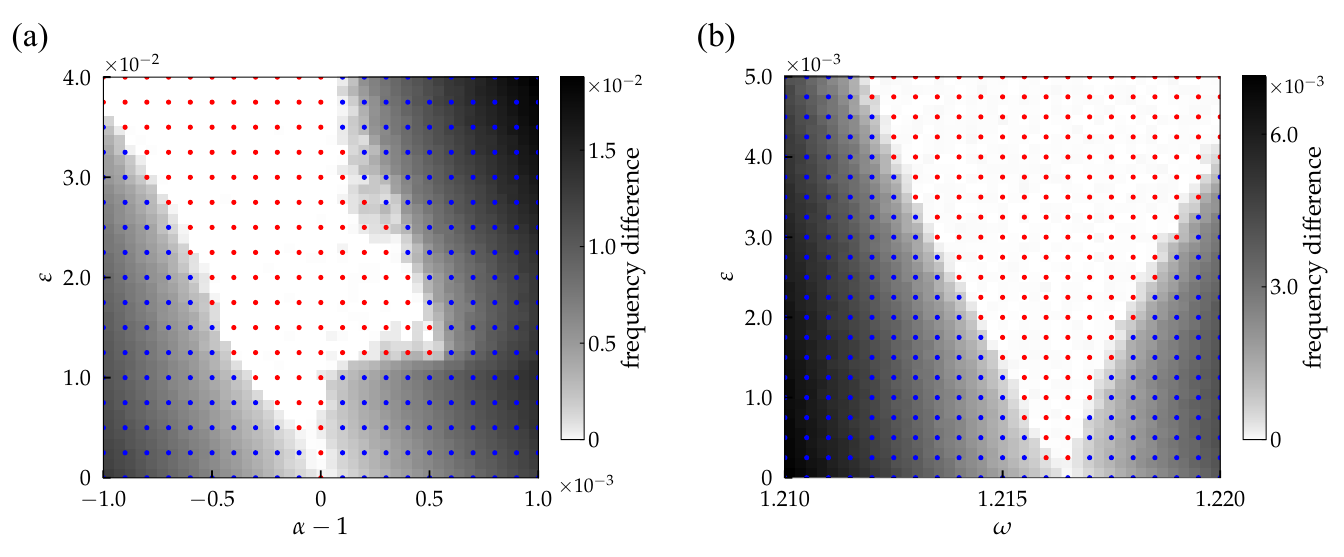}
    \caption{(a) The black-and-white heat map shows the frequency difference between the two oscillators in the system~\eqref{eq:double lorenz}. The red and blue dots represent the synchronous and asynchronous predictions by Eq.~\eqref{eq: main_result}, with red indicating synchronization and blue indicating asynchronization.
    (b) The same as (a), but for the system~\eqref{eq:rossler sprottn}.}
    \label{sfig:pred tongue}
\end{figure*}

We verify the validity of our theory not only for the parameters presented in the main text but also over a broader range.
Specifically, we systematically examined whether the oscillators are synchronized---namely, whether a phase difference $\psi$ exists such that the right-hand side of Eq.~\eqref{eq: main_result} in the main text vanishes---by evaluating Eq.~\eqref{eq: main_result} at all parameter points indicated by the red and blue dots in Fig.~\ref{sfig:pred tongue}. 
Red dots denote synchronized parameter sets, whereas blue dots represent asynchronous ones. 
As shown in Fig.~\ref{sfig:pred tongue}, the red and blue dots accurately predict the Arnold tongue, depicted by the white region in the background. 
The Arnold tongue of chaotic oscillators can take on anomalous shapes owing to qualitative changes in the oscillator trajectories induced by coupling, as illustrated in Fig.~\ref{sfig:pred tongue}(a). 
Our theory provides a reduction that faithfully captures even such unconventional synchronization characteristics of chaotic oscillators.

\subsection{Inferring Coupling Functions}
In this section, we provide implementation details for statistically inferring coupling functions from time-series data of chaotic oscillators.

We first describe the procedure used to generate the data for inference.  
From the attractors of the two oscillators composing the coupled chaotic systems in Eqs.~\eqref{eq:double lorenz} and~\eqref{eq:rossler sprottn}, we randomly select $5{,}000$ points as initial conditions.  
Numerical simulations are then performed from each initial condition using a fourth-order Runge--Kutta scheme with time step $h$, over the interval $t \in [0, T_{\mathrm{end}}]$ in arbitrary time units.  
The resulting time series are converted to phase using the cross sections described in the main text.  
The time derivative of the phase is computed numerically via finite differences, and both the phase and its derivative are sampled at a rate $f$.  
For the system in Eq.~\eqref{eq:double lorenz}, we set $h = 10^{-3}$, $T_{\mathrm{end}} = 300$, and $f = 0.4^{-1}$.  
For the system in Eq.~\eqref{eq:rossler sprottn}, we use $h = 10^{-2}$, $T_{\mathrm{end}} = 100$, and $f = 4.0^{-1}$.  

Next, we describe the details of the inference procedure using Gaussian process regression (GPR).
We emphasize that the choice of inference method is not essential; similar results can be obtained using other approaches such as Bayesian linear regression. 

In GPR, the characteristics of the function being inferred can be incorporated into the design of the kernel function beforehand.
In our case, because the coupling functions are $2\pi$-periodic, we adopt a periodic kernel defined as
\[
k(\psi, \psi') = \sigma_f^2 \exp \left[-\frac{2\sin^2 \left((\psi - \psi')/2\right)}{\ell^2}\right],
\]
where $\sigma_f$ and $\ell$ denote the output variance and length scale, respectively.
We set $\ell$ to 0.3 throughout the inference.
The mean function is chosen to be constant. 
To reduce computational cost, we employ the FITC (Fully Independent Training Conditional) approximation~\cite{Snelson2005} implemented as \texttt{gpflow.models.GPRFITC}. 
A total of $M=100$ fixed inducing points are uniformly distributed over $[0,2\pi]$.

Hyperparameters are optimized by maximizing the log marginal likelihood using the \texttt{gpflow.optimizers.Scipy()} routine. 
After optimization, the predictive mean and variance of the coupling function are evaluated at $1{,}000$ evenly spaced points over $[0,2\pi]$. 
All computations are implemented in GPflow~\cite{GPflow}.

\end{document}